
\headline={\ifnum\pageno=1\firstheadline\else
\ifodd\pageno\rightheadline \else\leftheadline\fi\fi}
\def\firstheadline{\hfil}
\def\rightheadline{\hfil}
\def\leftheadline{\hfil}
        \footline={\ifnum\pageno=1\firstfootline\else\otherfootline\fi}
\def\firstfootline{\rm\hss\folio\hss}
\def\otherfootline{\hfil}
\def\ick{\eqalignno}

\font\tenrm=cmr10

\font\elevenbf=cmbx10 scaled\magstep 1
\font\elevenrm=cmr10 scaled\magstep 1
\font\elevenit=cmti10 scaled\magstep 1

\font\ninerm=cmr9

\nopagenumbers
\hsize=6.0truein
\vsize=8.5truein
\parindent=1.5pc
\baselineskip=10pt
\line{\elevenrm\hfill UdeM-LPN-TH-94-202}
\vglue 5pt
\line{\elevenrm\hfill hep-th/9407030}
\vglue 1.0cm
\centerline{\elevenbf  RADIATIVE CORRECTIONS IN }
\vglue 7pt
\centerline{ {\elevenbf NONRELATIVISTIC CHERN-SIMONS THEORY }
\footnote{\hbox {$^*$}} { \ninerm\baselineskip=11pt
Talk given by M.L. at the MRST94 Conference, ``What Next? Exploring the
Future of High Energy
Particle Physics", McGill University, Montreal, Canada, May 1994.
\hfil} }

\vglue 7pt
\vglue 1.0cm
\centerline{\elevenrm DIDIER CAENEPEEL $^{1}$  and MARTIN LEBLANC $^{2}$,}
\baselineskip=13pt
\centerline{\elevenit $^{1}$ Laboratoire de Physique Nucl\'eaire,}
\baselineskip=12pt
\centerline{\elevenit $^{2}$ Centre de Recherches Math\'ematiques, }
\baselineskip=12pt
\centerline{\elevenit Universit\'e de Montr\'eal, Montr\'eal, Qc, H3C-3J7 }

\vglue 0.8cm
\centerline{\tenrm ABSTRACT}
\vglue 0.3cm
  {\rightskip=3pc
 \leftskip=3pc
 \tenrm\baselineskip=12pt
 \noindent
We present the one-loop
scalar field effective potential for the $N=2$ supersymmetric
nonrelativistic self-interacting matter fields coupled to an Abelian
Chern-Simons gauge field and for its generalization when bosonic
matter fields are coupled to non-Abelian
Chern-Simons field. In both models, Gauss's law linearly relates the
magnetic field to the matter field densities;
hence, we also include radiative effects from the
background gauge field. We compute the scalar field effective potentials in two
gauge families, a gauge reminiscent of the $R_\xi$-gauge in the limit
$\xi\rightarrow 0$ and in the Coulomb family gauges. We regularize the theory
with operator regularization and a cutoff to demonstrate that the results are
independent of the regularization scheme.
\vglue 0.8cm }
\line{\elevenbf 1. The models.\hfil}
\vglue 0.2cm
\line{\elevenit 1.1. $N=2$ supersymmetric nonrelativistic Chern-Simons
model \hfil}
\vglue 0.1cm
\baselineskip=14pt
\elevenrm

The first model we consider
is composed of self-interacting scalar field and fermionic
field coupled to an Abelian Chern-Simons gauge field $^1$
[${\rm diag}\;\;\eta~=(+,-,-)$]
$$\ick {
S= \int dt d^2{\bf x} \;\;
&\Bigl \{{\kappa\over 2} (\partial_t {\bf A}) \times {\bf A} -\kappa
A^0 {\bf \nabla} \times {\bf A}
+i\phi^*(\partial_t +i A^0 )\phi +i\psi^*(\partial_t +iA^0)\psi \cr
& - {1 \over 2 } |{\bf D}\phi|^2  - {1 \over 2 } |{\bf D}\psi|^2
+ {1\over 2} B |\psi|^2 -{\lambda_1 \over 4}(|\phi|^2)^2
-\lambda_2|\phi|^2 |\psi|^2 \Bigr \}&(1.1) \cr }
$$
where ${\bf D}={\bf \nabla }-i{\bf A}$ is the covariant derivative
and $B={\bf \nabla}\times {\bf A}$ is the magnetic field.
Note that the fermionic field is
non-minimally coupled to the gauge field through the Pauli term.
We have omitted the mass parameter since in
nonrelativistic systems, it is always possible to set it equal to unity.
[We use
a vector notation: in the plane the cross product
is ${\bf V}\times {\bf W} = \epsilon^{ij}V^iW^j$, the curl of a vector is
${\bf \nabla}\times {\bf V}=\epsilon^{ij}\partial_iV^j$, the curl of
a scalar is $({\bf\nabla}\times S)^i=\epsilon^{ij}\partial_jS$ and we shall
introduce the notation $\bigl ({\bf A}\times {\bf {\hat z}}\bigr)^i
=\epsilon^{ij}A^j$.
The notation $x=(t, {\bf x})$ will also be used unless stated otherwise.]

The system (1.1) enjoys several invariances at the classical level.
The system is Galilean, conformal, gauge invariant $^{2,3}$ and
$N=2$ supersymmetric when $\lambda_1= -{2\over \kappa}$
and $\lambda_2={3\over 4}\lambda_1$ $^1$. At those values of
the coupling constants, the model admits static self-dual (soliton)
solutions $^3$.

\vglue 0.4cm
\line{\elevenit 1.2. Nonrelativisitic non-Abelian Chern-Simons model.\hfil}
\vglue 0.1cm
The second model we consider is
a self-interacting scalar field coupled to SU(2) non-Abelian Chern-Simons gauge
fields, which generalizes the above system
$$\eqalignno {
S= \int dt d^2{\bf x} \;\;
&\Bigl \{- \kappa \epsilon^{\alpha\beta\gamma}{\rm Tr}
(A_\alpha \partial_\beta A_\gamma + {2\over 3}A_\alpha A_\beta A_\gamma)
+i\phi^\dagger D_t \phi
- {1 \over 2 } |{\bf D}\phi|^2
-{\lambda_{pqrs}\over 4}\phi^\dagger_p\phi^\dagger_q \phi_r \phi_s
\Bigr \} \quad
&(1.2) \cr }
$$
where the gauge fields belong to the su(2) Lie algebra
$A_\mu=i{A_\mu^a \tau^a\over 2} $,
and $ D_t = \partial_t + i A_0^a {\tau^a\over 2}$ and
${\bf D}={\bf \nabla }-i{\bf A}^a {\tau^a\over 2}$
are the time and space covariant derivatives respectively. $\phi_p$ is the
two component nonrelativistic scalar field, $p=1,2$. The self-interaction
coupling constants satisfy
$\lambda_{pqrs}= \lambda_{qpsr}$ since the fields are bosonic
and $\lambda^*_{pqrs}=\lambda_{rspq}$ for the Lagrangian to be real.
$\tau^a$ are the Pauli matrices
which satisfy the usual commutation relations
$[{\tau^a\over 2},{\tau^b\over 2}]=\epsilon^{abc}{\tau^c\over 2}$
and trace relation ${\rm Tr} \bigl( {\tau^a\over 2}{\tau^b\over 2}\bigr)
= {1\over 2} \delta^{ab}$.

Again, the system (1.2) enjoys the
Galilean and conformal symmetries but we speak of ``gauge invariance"
only when a special quantization condition holds
for the Chern-Simons coupling constant $4\pi\kappa={\rm const.}$ $^4$
and if
$$
\lambda_{1111}=2\lambda_{1212}=\lambda_{2222} \equiv \lambda \quad.\eqno (1.3)
$$
As above, self-dual (soliton) solutions exist in this model $^5$.

Our goal is to compute the scalar field effective potentials for both
nonrelativistic Chern-Simons matter systems using a functional method $^{6,7}$.

\vglue 0.6cm
\line{\elevenbf 2. Scalar field effective potential for the Abelian
model.\hfil}
\vglue 0.4cm
One can ask whether any of the classical symmetries survive the
quantization of the theory. Furthermore, one can
wonder if the self-dual solutions are stable or
even if there are any modifications to the form of the potential.
Many methods have been constructed to assess those effects.
For example, the well-know Feynman's diagrammatic
or the functional methods have been useful
in such studies. Here, we will concentrate on the latter method to compute
the effective potentials for both models.

The effective potential method starts with the definition of a new
shifted action~$^8$
$$\ick {
S_{\rm new}&=S\Bigl\{\phi(x)=\varphi+\pi(x);\psi(x);
A^\mu (x)=a^\mu(x)+Q^\mu(x)\Bigr \} &(2.1) \cr
&\quad -S\Bigl \{ \varphi,a^\mu(x)\Bigr\} - {\rm terms\;linear \;in \;quantum\;
fields,} \cr }
$$
where we shift the scalar field by a constant field and we shift
the gauge field by a solution to the classical equations of motion for the
electromagnetic fields.
To maintain consistency with Gauss's law, which relates linearly the
magnetic field to the matter field, we need to
choose a background gauge field $a^\mu(x)$ such that the magnetic field is
constant throughout the plane. We set ${\bf a}({\bf x})
= -{B\over 2}{\bf x}\times {\bf {\hat z}}
={\varphi^*\varphi\over 2\kappa}{\bf x}\times {\bf {\hat z}}$ where $B$ is the
constant magnetic field. Such a choice is also
consistent with the electric field equation of motion if
$a^0({\bf x})=-{(\varphi^*\varphi)^2\over 4\kappa^2}
{\bf x}^2$. The
fermionic field is not shifted because we consider only quantum corrections
to the scalar field effective potential.
Note however, that this solution for the background vector potential
and with a constant $\varphi$ does not provide a solution to the equation
for the matter sector unless $\varphi=0$.

Next, one chooses a gauge-fixing condition. We performed the calculation with
a Coulomb-gauge
$
{\cal L}_{G.F.}= {1\over 2\xi}({\bf \nabla}\cdot {\bf Q})^2
$ for arbitrary $\xi$ and with an $R_\xi$-gauge
$
{\cal L}_{G.F.}={1\over 2\xi}\bigl [ {\bf \nabla }\cdot{\bf Q}
+ i\xi\varphi\pi^*\bigr]
{}~\bigr [ {\bf \nabla }\cdot {\bf Q} - i\xi\varphi^*\pi\bigr]
$
(in the $\xi\to 0$ limit).
The second one turns out to be
the most efficient gauge-fixing condition since it cancels
unwanted cross terms. The shifted action then becomes ($R_\xi$-gauge)
$$\eqalignno {
S=\int dt & \; d^2{\bf x} \;
\Bigl\{ {\kappa\over 2}(\partial_t{\bf Q})\times {\bf Q}
-\kappa Q^0 {\bf \nabla}\times {\bf Q}
+ {1\over 2\xi}({\bf \nabla}\cdot {\bf Q})^2 - {\rho\over 2}{\bf Q}\cdot
{\bf Q} \cr
&+i\pi^*(\partial_t +ia^0)\pi -{1\over 2}|{\bf D}\pi|^2 - {\lambda_1\over 4}
\Bigl (\varphi^2 (\pi^*)^2+ 4 \rho |\pi|^2 + (\varphi^*)^2(\pi)^2
\Bigr )+{\xi\over 2}\rho |\pi|^2 \cr
&+i\psi^*(\partial_t +ia^0)\psi -{1\over 2}|{\bf D}\psi|^2 -
\lambda_2 \rho |\psi|^2
+{B \over 2}|\psi|^2 \cr
&+c^*(-\nabla^2 +\xi\rho )c
+J^0Q^0 + {\bf J}\cdot {\bf Q}\Bigr\} &(2.2)  \cr}
$$
with $\rho=\varphi^*\varphi$,
$J^0=-[\varphi^*\pi+\pi^*\varphi]$ and ${\bf J}={\bf a}J^0$.
The c-field term is the ghost compensating term arising from the choice
of gauge-fixing condition.
After a change of variable, we can write the shifted action in a form
readily integrable
$$\eqalignno {
\int dt dt' d^2{\bf x} d^2{\bf x'} \;
&\Bigl \{
{1\over 2} \pi^{*a}(x) {\cal D}^{-1}_{ab}(x-x') \pi^b(x')\cr
&+{1\over 2} \psi^{*a}(x) {\cal S}^{-1}_{ab}(x-x') \psi^b(x')
-{1\over 2}{Q^{\prime}}^\mu(x) {\Delta}^{-1}_{\mu\nu}(x-x')
{Q^{\prime}}^\nu (x')&(2.3)\cr
&+c^*(x){\cal P}^{-1}(x-x')c(x')
+ {1\over 2} J^\mu (x) {\Delta}_{\mu\nu} (x-x') J^\nu (x') \Bigr \}
\cr}
$$
Upon performing the path integrals, we find the effective action
to be
$$\eqalignno {
\Gamma_{\rm eff}&= S(\varphi,a^\mu({\bf x})) +
{i\over 2}\ln {\rm Det} \{{\cal D}^{-1}_{ab} + {\cal M}_{ab}\}
-i \ln {\rm Det} {\cal S}^{-1}_{ab}
+{i\over 2} \ln {\rm Det} \Delta^{-1}_{\mu\nu} - i \ln {\rm Det} {\cal P}^{-1}
&(2.4) \cr }
$$
where the matrices are easily found from Eq.(2.2) and the presence of the
${\cal M}_{ab}$ matrix is due to the mixing between matter and gauge fields.

We now discuss the structure of the perturbative expansion.
The effective potential is related to the effective action by
$V_{\rm eff}\int d^3x=-\Gamma_{\rm eff}$ when
defined on constant background fields. In the present case, the background
gauge field $a^\mu({\bf x})$
is space-dependent; hence, we cannot use directly the functional
method of Jackiw.
We adopt the following strategy. We will compute the effective action by
factoring out a matrix that is
background gauge field independent and perturbatively expand the gauge field
dependent part in powers of small coupling constants $\lambda_1\ll 1$,
$\kappa^{-1}\ll 1$ (recall that $a^0\sim {\rho^2\over
\kappa^2}$ and ${\bf a }\sim {\rho\over \kappa}$).
The computation is up to ${\cal O}(\rho^3)$ because each term of
${\cal O}(\rho^3)$ is either of ${\cal O}(\lambda_1^3)$,
${\cal O}({\lambda_1^2\over \kappa})$ or ${\cal O}({1\over \kappa^3})$.
Therefore, for the rest of the paper, we will use the terminology
${\cal O}(\rho^3)$.
We do not introduce the parameter $\lambda_2$ in these expressions for
simplicity since
$\lambda_2$ does not enter the scalar field effective potential, as we
will see below. Thus, as again we will see, we need only to consider 1-point or
2-point functions in background gauge fields that contribute to the effective
action. The gauge-independent part will be treated following Jackiw's
method.
We will also take the special limit over the gauge parameter
$\xi\to 0$. In this limit $\ln det \Delta^{-1} = 0$
and $\ln det{\cal P}^{-1} = 0$.

The contributions from the gauge field $a_\mu (x)$ can be obtained
by factorization, for example,
$
\ln {\rm Det}\bigl( {\cal D}^{-1}+{\cal M}\bigr )
= {\rm Tr}{\rm Ln}\Bigl( \Theta^{-1} \bigl ( 1
+ \Theta X \bigr )\Bigr )
$ and then the
${\rm Ln} (1+\Theta X)$ can be expanded in power of $X$, which is gauge field
dependent. Upon doing this, we find that all contribution
from $a_\mu$ vanishes to the order considered.
Thus, we are left only with matter contributions
$$\eqalignno {
V_{\rm eff} = -{\Gamma_{\rm eff}\over \int d^3x} &= V_0(\rho)
-{i \over 2}\int {d^2{\bf p} \over (2\pi)^2} {d\omega \over 2\pi}\ln
{1\over \mu'^4}
\Bigl \{ -\omega^2 + \bigl ({{\bf p}^2 \over 2} +\lambda_1\rho\bigr )^2
+ \bigl (
-{\lambda_1^2 \over 4}+ {1 \over \kappa^2} \Bigr )\rho^2
+{\cal O}(\rho^3)
\Bigr \} \cr
&+i \int {d^2{\bf p} \over (2\pi)^2} {d\omega \over 2\pi}\ln
{1\over \mu'^4}
\Bigl \{ -\omega^2 + \bigl ({{\bf p}^2 \over 2} +\lambda_2\rho-{1\over 2}B
\bigr )^2  \Bigr \} &(2.5)
\cr  }
$$
where the first integral comes from the bosonic integration, while
the second integral comes from the fermionic integration.

We pose for a moment to notice that so far, no regulator has been used.
In fact, divergences occurs only in the matter sector. We have performed the
regularization of the expression in Eq.(2.5) with a cut-off and
operator regularization $^6$. The result
is the same whether one uses the former or the latter.
We present here the result using operator regularization. The basic ingredient
is the regularization of the logarithm
$$\eqalignno {
{\rm Tr} \ln H&=-\lim_{s\to0}{d \over ds}
{\rm Tr} {1\over \Gamma(s)}\int_0^\infty dt\; t^{s-1}
\Bigl\{e^{-H_0t}+e^{-H_0t}(-t)H_I+e^{-H_0t}{(-t)^2\over 2}H_I^2+\dots\Bigr\}
&(2.6) \cr}
$$

Upon making the identification
$H_0=\bigl
[-\omega ^2+ \bigl ({{\bf p}^2 \over 2}+\lambda_1\rho\bigr )^2\bigr ]/\mu'^4
$
and
$H_I~=\Bigl(-{\lambda_1^2 \over 4} + {1 \over \kappa^2} \Bigr )
\rho^2/\mu'^4
$ for the boson, and
$H_0= \Bigl
[-\omega^2 + \bigl ({{\bf p}^2 \over 2} +\lambda_2\rho-{1\over 2}B
\bigr )^2\Bigr ]/\mu'^4
$ for the fermion, dropping an unimportant ${\rm const.}\rho^2$-term
coming from the first term in Eq.~(2.6), computing the one-pts function,
evaluating the energy/momentum integrals, and
imposing the normalization condition
$
{d^2 \over d \rho^2}V_{\rm eff}|_{\rho=\mu^2}={1 \over 2}\lambda_1(\mu),
$
we find to ${\cal O}(\rho^3)$ in the $\xi\to 0$ limit the result
$$
\eqalignno {
V_{\rm eff}(\rho )=&{1 \over 4}\lambda_1(\mu)\rho^2
+{\hbar \over 32\pi}\bigl (\lambda_1(\mu)^2-{4 \over \kappa^2}
\bigr )\rho^2 \bigl (\ln{\rho \over \mu^2}-{3 \over 2}\bigr ) \quad .&(2.7)\cr
}
$$

\vglue 0.6cm
\line{\elevenbf 3. Scalar field effective potential for the non-Abelian model.
\hfil}
\vglue 0.4cm
We follow the same procedure used in the above section to compute the
effective potential in the non-Abelian model. In the action (1.2),
we shift the scalar field by a constant but this time the gauge field
also can be shifted by a constant since the classical equation of motion
for the electromagnetic field can be solved by a constant gauge field $^9$.
In fact, for $\lambda=-5/16\kappa$ even the equation of motion for the
scalar field is satisfied without having to resort to a vanishing scalar
field. In such a choice of background gauge field, only global gauge
invariance is retained and again it is sufficient to gauge-fix the
action with ${\cal L}_{G.F.} = (1/2\xi)
(\nabla\cdot{\bf Q^a}+{i\over 2}\xi
\pi^\dagger \tau^a \varphi)
(\nabla\cdot{\bf Q^a}-{i\over 2}\xi \varphi^\dagger \tau^a \pi )$. After
writing the shifted action, adding the gauge-fixing term and compensating
ghosts, one can integrate over the three gauge fields one after the other
and ghost fields
easily, since the form of the operators in the determinants are diagonal in
Fourier space. We find that the ghost contribution cancels against
one contribution coming from the gauge field integration, which
is the pure background gauge field dependent part. The other contribution
coming from the gauge field integration modifies the matter sector.
Again, performing the integration over the matter fields and
regulating via operator regularization (or a cut-off), we find
to ${\cal O}(\rho^3)$ in the $\xi\to 0$ limit
and after (re)normalization
$$\eqalignno{
V_{\rm eff}(\rho,a_\mu^a)&={1\over 4}\rho^2\lambda(\mu)
+{\hbar\over 32\pi}
\bigl(\lambda^2(\mu)-{4\over \kappa^2}{3\over 16}\bigr)\rho^2
\bigl( \ln{\rho \over\mu^2} - {3\over 2}\bigr ).&(3.1)\cr }
$$
\vglue 0.6cm
\line{\elevenbf 4. Summary \hfil}
\vglue 0.4cm
We have computed the scalar field effective potential for matter fields coupled
to Abelian or non-Abelian Chern-Simons gauge field including radiative
corrections from a background gauge field consistent with Gauss's law. In both
models, we choose to gauge-fix with an $R_\xi$-gauge in the limit $\xi\to 0$ or
with a Coulomb gauge with arbitrary $\xi$ and we regulate divergences of the
matter sector either with operator regularization or with a cut-off. We find
that the answer is independent of the gauge-fixing condition and independent of
the regulator used. Our results agree with the ones found in the literature
$^{10-13}$.

Note that no ultraviolet divergences occur in the course of the evaluation of
the effective potentials as it is expected from using operator regularization;
however, when a cut-off is used we need to renormalize the theory with adequate
counter-terms. Our results Eq.~(2.7) and Eq.~(3.1) are independent of the
background gauge fields and the fermions do not contribute to the effective
potential in the Abelian model.
As a spin-off of our calculation, we find no infinite nor finite
renormalization of the Chern-Simons coupling constant in both models $^7$.

Finally, both models experiences radiative conformal symmetry-breaking for
unrelated coupling constants. The $\beta$-function can be read from the
renormalization group equation and we find
$$\eqalignno {&
\beta(\lambda(\mu))={1 \over 4\pi}\Bigl (\lambda^2(\mu)-{4 \over \kappa^2}
\alpha^2
\Bigr )&(4.1)\cr}
$$
where the subscript 1 has been dropped for the Abelian case and the group
factor $\alpha^2$ is 1 for the Abelian model and $3/16$ for the non-Abelian
model. Notice that in both models the conformal symmetry is restored upon
choosing the self-dual critical~point.
\vfill\eject
\vglue 0.6cm
\line{\elevenbf Acknowledgements \hfil}
\vglue 0.4cm
We thank F. Gingras and D.G.C. McKeon for their collaboration and
the Natural Sciences and Engineering Research Council of Canada and
the Fonds pour la Formation de Chercheurs et l'aide \`a la Recherche
for financial support.
\vglue 0.6cm
\line{\elevenbf References \hfil}
\vglue 0.4cm

\medskip
\item{1.} G. Lozano, M. Leblanc and H. Min, Ann. of Phys.{\bf 219}, 328 (1992).
\medskip
\item{2.} C.R. Hagen, Phys. Rev. {\bf D31}, 848 (1985).
\medskip
\item{3.} R. Jackiw and S.Y. Pi, Phys. Rev. {\bf D49}, 3500 (1990).
\medskip
\item{4.} S. Deser, R. Jackiw and S. Templeton, Phys. Rev. Lett.
{\bf 48}, 975 (1982); Ann. Phys. (N.Y.) {\bf 140}, 372 (1982).
\medskip
\item {5.} G. Dunne, R. Jackiw, S. Pi and C. Trugenberger,
Phys. Rev. {\bf D 43}, 1332 (1991).
\medskip
\item {6.} D. Caenepeel, F. Gingras, M. Leblanc and D.G.C. McKeon,
Phys. Rev. {\bf D49}, (1994) (in press).
\medskip
\item {7.} D. Caenepeel and M. Leblanc, preprint UdeM-LPN-TH-94-200, (1994).
\medskip
\item {8.} R. Jackiw, Phys. Rev. {\bf D9}, 1686 (1974).
\medskip
\item {9.} L.S. Brown and W.I. Weisberger, Nucl. Phys. {\bf B157}, 285 (1979).
\medskip
\item {10.} G. Lozano, Phys. Lett. {\bf B283}, 70 (1992).
\medskip
\item {11.} O. Bergman and G. Lozano, Ann. Phys. (N.Y.) {\bf 229}, 416 (1994).
\medskip
\item {12.} D. Freedman, G. Lozano and N. Rius, Phys. Rev. {\bf D49}, 1054
(1994).
\medskip
\item {13.} D. Bak and O. Bergman, preprint MIT-CTP-2283, (1994).
\medskip

\vfill\eject
\bye